\documentclass[%
 preprint,
superscriptaddress,
 amsmath,amssymb,
 aps,
 prl,
]{revtex4-1}

\usepackage[utf8]{inputenc}
\usepackage{booktabs}
\usepackage{graphicx}
\usepackage[table,xcdraw]{xcolor}
\usepackage{float}

\usepackage{amssymb, amsmath} 
\usepackage[citecolor=black]{hyperref}
\usepackage[margin=0.5in]{geometry}

\usepackage[table,xcdraw]{xcolor}
\usepackage{hyperref}
\usepackage{cleveref}
\newcommand*\chem[1]{\ensuremath{\mathrm{#1}}}
\usepackage[center]{caption}

\begin{document}

\title{Combining time-dependent density functional theory and the $\Delta$SCF approach for accurate core-electron spectra}

\author{Marcus Annegarn}
\affiliation{Departments of Materials, Imperial College London, London SW7 2AZ, United Kingdom}
\affiliation{The Thomas Young Centre for Theory and Simulation of
Materials, Imperial College London, London SW7 2AZ, United Kingdom}

\author{Juhan Matthias Kahk}
\affiliation{Institute of Physics, University of Tartu, W. Ostwaldi 1, 50411 Tartu, Estonia.}

\author{Johannes Lischner}
\affiliation{Departments of Materials, Imperial College London, London SW7 2AZ, United Kingdom}
\affiliation{The Thomas Young Centre for Theory and Simulation of
Materials, Imperial College London, London SW7 2AZ, United Kingdom}

\begin{abstract}
Spectroscopies that probe electronic excitations from core levels into unoccupied orbitals, such as X-ray absorption spectroscopy and electron energy loss spectroscopy, are widely used to gain insight into the electronic and chemical structure of materials. To support the interpretation of experimental spectra, we assess the performance of a first-principles approach that combines linear-response time-dependent density (TDDFT) functional theory with the $\Delta$~self\nobreakdash-consistent field ($\Delta$SCF) approach. In particular, we first use TDDFT to calculate the core-level spectrum and then shift the spectrum such that the lowest excitation energy from TDDFT agrees with that from $\Delta$SCF. We apply this method to several small molecules and find encouraging agreement between calculated and measured spectra.
\end{abstract}

\maketitle

\section{Introduction}
X-ray absorption spectroscopy (XAS) and electron energy loss spectroscopy (EELS) are powerful and widely used characterization techniques that can provide information about the elements present in a sample as well as their chemical environments. For example, these techniques have been used for studying the electronic structure of functional materials~\cite{doi:10.1063/1.4984072,Abbehausen2018a}, probing the chemical bonding in systems such as water and ice~\cite{Zhovtobriukh2019a, Cavalleri2002a} or analyzing the properties of pollution particles~\cite{Datta2012a,Chen2005b}.

Both XAS and EELS measure energies of electronic excitations from core levels into unoccupied states. The onset of the spectrum corresponding to transitions from 1s core states into the lowest unoccupied orbitals is called the K-edge, while the L-edge indicates the onset of transitions from core states with a principal quantum number of $n=2$. While it is usually straightforward to use XAS and EELS for elemental analysis of samples, the identification of specific chemical environments can be more challenging. In principle, assignment of chemical environments should be possible by comparing the measured spectrum to a set of experimental reference spectra. In practice, however, obtaining reliable reference spectra for a broad range of chemical environments is often not straightforward.

Alternatively, it is possible to obtain reference spectra from first principles calculations. For example, linear response time-dependent density functional theory (TDDFT)~\cite{casida1995time} has been widely used to predict core-excitation energies and intensities of XAS spectra~\cite{besley2010time,oosterbaan2018non,besley2007time,norman2018simulating}. Core spectra from TDDFT often reproduce accurately the qualitative shape of measured spectra, but absolute core-level excitation energies are not quantitatively reproduced with errors of several electron volts~\cite{wenzel2014calculating,attar2017femtosecond,bhattacherjee2018photoinduced} (unless one employs short-ranged corrected functionals trained on core excitations~\cite{besley2009time}). Often good quantitative agreement with experiment can be obtained by subtracting a constant energy shift from all excitation energies with the value of the shift being determined empirically by comparison to experiment~\cite{debeer2010calibration}. 

Accurate absolute core-level excitation energies have recently been obtained using the $\Delta$ Self-Consistent-Field ($\Delta$SCF) approach~\cite{gilbert2008self,kowalczyk2011assessment,ziegler1977calculation} in which the excitation energy is determined as a total energy difference. For transitions from core orbitals to the vacuum level which are probed in X-ray photoemission experiments, some of us recently demonstrated that highly accurate excitation energies (also known as core-level binding energies) can be obtained for molecules, solids and surfaces when the SCAN exchange-correlation functional is used in conjunction with an accurate treatment of relativistic effects~\cite{Kahk2018a,Kahk2019b}. The $\Delta$SCF approach has also been used to calculate K-edge energies and energies of higher-lying excited states~\cite{Ambroise2019a,Zhekova2014a}. For the latter, convergence difficulties associated with a variational collapse are often encountered~\cite{Zhekova2014a}. To overcome this problem, Hait and Head-Gordon used a square gradient minimization approach and obtained good agreement with experiment for a set of molecular compounds~\cite{Hait2019b,Hait2020a}. However, the determination of spectra with this approach is less straightforward as a separate calculation is required for each excited state (in contrast to linear-response TDDFT which yields all excitation states in a single shot).

Other approaches for simulating core-electron spectra are based on Kohn-Sham eigenvalues~\cite{Tait2016a,morris2014optados}, the coupled cluster approach~\cite{Coriani2012} and the GW/BSE approach~\cite{kehry2020,Gilmore2015a}. Finally, machine learning techniques have been developed to predict spectra of complex materials, but these techniques also need accurate reference spectra to construct the training data set~\cite{Zheng2018a,Torrisi2020a,Aarva2019a,Rankine2020}.

A simple alternative approach to obtain core-level spectra is to combine $\Delta$SCF approach with linear response TDDFT~\cite{leetmaa2006recent}: instead of using experimental data to determine the energy shift that is applied to the TDDFT spectrum, one determines this shift from first principles using a $\Delta$SCF calculation of the lowest excited state. In this paper, we apply this approach to a set of molecules and assess its accuracy by comparing the calculated spectra to experimental results. For most systems, we find good quantitative agreement when appropriate exchange-correlation functionals are used. It is straightforward to apply this method to more complex systems, such as surfaces or clusters.

\section{Methods}

The core-level spectra are obtained from
\begin{equation}
    S(\omega) = \sum_I f_I L^{(\eta)}(\hbar \omega - E_I),
\end{equation}
with $\omega$ denoting the light frequency, $f_I$ is the oscillator strength of the $I$-th excited state and $L^{(\eta)}$ denotes a Lorentzian with a full width ad half maximum (FWHM) of $\eta$. Also, $E_I$ is the energy of the $I$-th excited state and obtained via
\begin{equation}
    E_I = E_I^{\text{TDDFT}} +\Delta,
\end{equation}      
where $E^{\text{TDDFT}}_I$ are the excitation energies obtained from TDDFT and $\Delta$ denotes a constant energy shift that is applied to all excitation energies. The shift is given by
\begin{equation}
    \Delta =  E^{\Delta \text{SCF}}_1 - E_1^{\text{TDDFT}},
\end{equation}
with $I=1$ referring to the lowest excitation and $E^{\Delta \text{SCF}}_1$ denotes the corresponding excitation energy obtained from $\Delta$SCF.

For example, Table~\ref{table1} shows the lowest neutral excitation energy of NH$_3$ calculated with TDDFT using different exchange-correlation functionals (BHLYP, BLYP, PBE0 and Hartree-Fock (HF)). The calculated results differ from the experimental value at least by several eV. In contrast, the $\Delta$SCF results for all exchange-correlation functionals are in very good agreement with experiment. 

\begin{table}[h!]
\centering
\begin{tabular}{|c || c | c | c | c | c |} 
 \hline
  & BHLYP (eV) & BLYP (eV) & PBE0 (eV) & HF (eV) & exp. (eV)  \\
 \hline
 TDDFT & 398.7 & 379.9 & 389.1 & 416.0 & 400.4 \\
 $\Delta$SCF & 400.4 & 400.5 & 400.6 & 400.5 & 400.4\\
  \hline
\end{tabular}
\caption{Comparison of K-edge energies of NH$_3$ from TDDFT and the $\Delta$SCF approach for different exchange-correlation functionals. The experimental result is taken from Ref.~\cite{Nilsson2010b}
\label{table1}}
\end{table}

In the following we describe in more detail how the $\Delta$SCF calculations and the linear-response TDDFT calculations are carried out. For all molecules, calculations were carried out for the relaxed structure obtained using the SCAN functional with the default tight basis sets in the FHI-AIMS computer programme ~\cite{Blum2009,Ren2012}.

\subsection{$\Delta$SCF approach}

In the $\Delta$SCF approach, excitation energies are obtained as total energy differences between the relevant excited states and the ground state. For example, core-electron binding energies can be obtained by subtracting the ground state energy of the neutral system from the total energy of the system with a core hole which is obtained by minimizing the total energy under the constraint that a given core orbital remains unoccupied. Similarly, the lowest neutral core-electron excitation energy can be obtained by calculating the total energy of the system with a core hole and an extra electron in the lowest unoccupied orbital and subtracting this from the ground state energy (without a core hole).

The $\Delta$SCF calculations were performed using the FHI-AIMS computer programme~\cite{Blum2009,Ren2012}, an all-electron code that uses localized basis sets defined on a grid of points in real space. We include relativistic effects using the scaled Zeroth Order Regular Approximation (ZORA) \cite{doi:10.1063/1.467943,FAAS1995632,doi:10.1063/1.479395,doi:10.1063/1.1323266, Blum2009}. All molecules in this study are closed-shell, but spin polarization is included in the calculations with a core hole. It has been pointed out that the $\Delta$SCF approach cannot properly model singlet excitations as those cannot be described by a single Slater determinant~\cite{Hait2019b}. Despite this shortcoming, we find below that the $\Delta$SCF approach produces accurate K-edge energies - likely because the coupling between the core hole and the excited electron is weak. 

The basis sets used are those described in Refs.~ \cite{Kahk2018a,Kahk2019b}, which are modifications of the default tight FHI-AIMS basis sets with additional basis functions for the core states. These additional functions allow us to describe the contraction of the occupied 1s state in the presence of a core hole. The following five different exchange-correlations functionals were tested: SCAN, BHLYP, BLYP, PBE0, PBE, and B3LYP. We stress that the same computational parameters (such as basis set and exchange-correlation functional) must be used in both the ground state and the excited state calculation to obtain accurate excitation energies. Particular care must be taken when the molecules contain atoms in equivalent sites. In this case, additional strategies for localizing the core holes on a desired atom must be employed as explained in~\cite{Kahk2019b}.

\subsection{Time-dependent density-functional theory}

We also carry out linear-response TDDFT calculations of excited states involving transitions from core orbitals to unoccupied states using the NWChem programme package~\cite{doi:10.1063/5.0004997}. For this, we ignore (or "freeze") transitions from all occupied states with energies exceeding that of the core orbital under consideration. We have verified that inclusion of these states only leads to very small changes in the excitation energies.

\begin{figure}[h]
    \centering
    \includegraphics[scale=0.75]{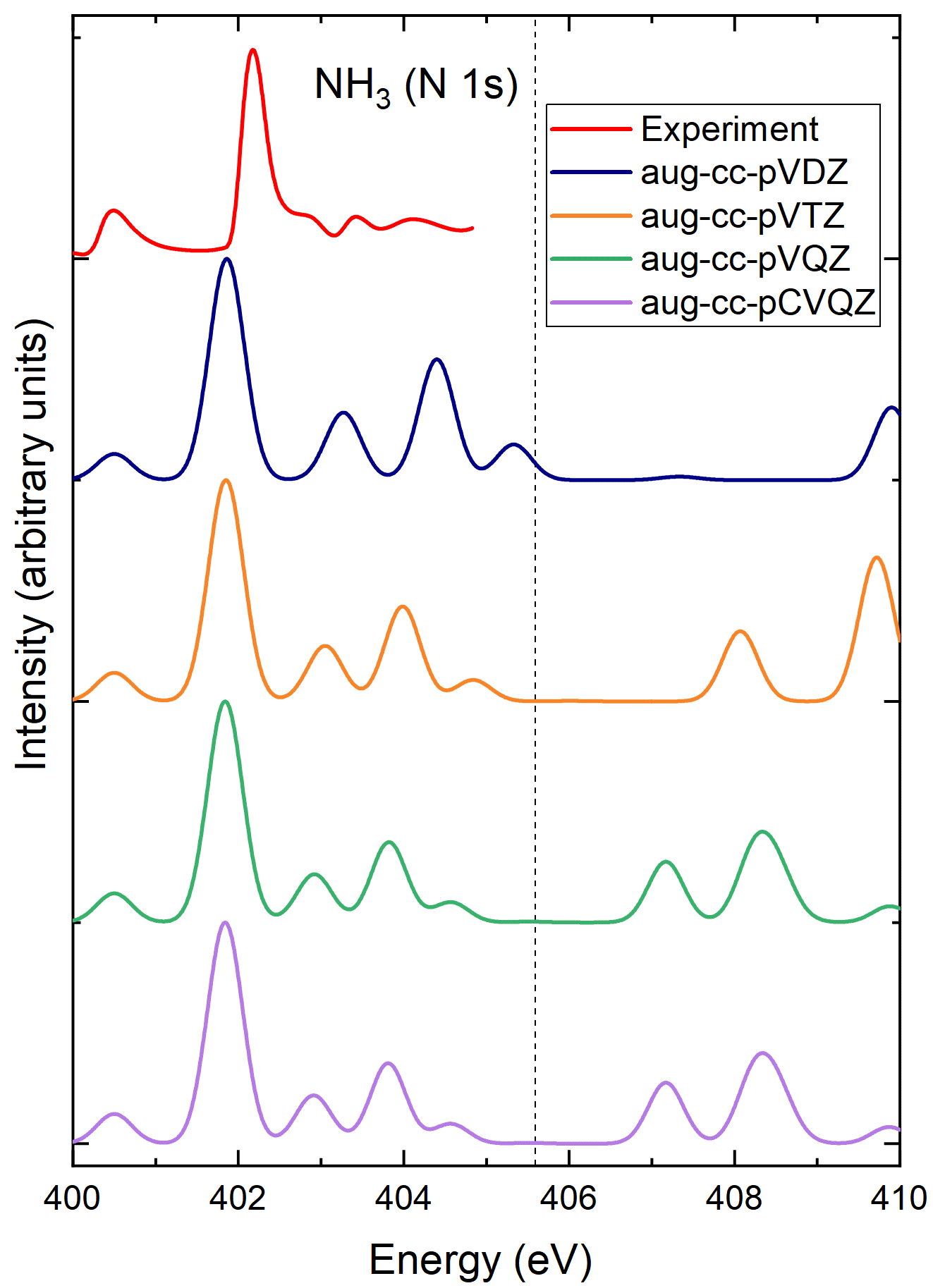}
    \caption{Comparison of measured XAS spectrum from Ref.~\cite{Nilsson2010b} and TDDFT results with basis sets for NH$_3$. The calculated spectra were obtained with the BHLYP functional and shifted such that the energy of the first excited state from TDDFT agrees with the BHLYP $\Delta$SCF result. The dashed vertical line indicated the calculated ionization potential.}
    \label{basis_comp}
\end{figure}

\begin{figure}[h]
    \centering
    \includegraphics[scale=0.75]{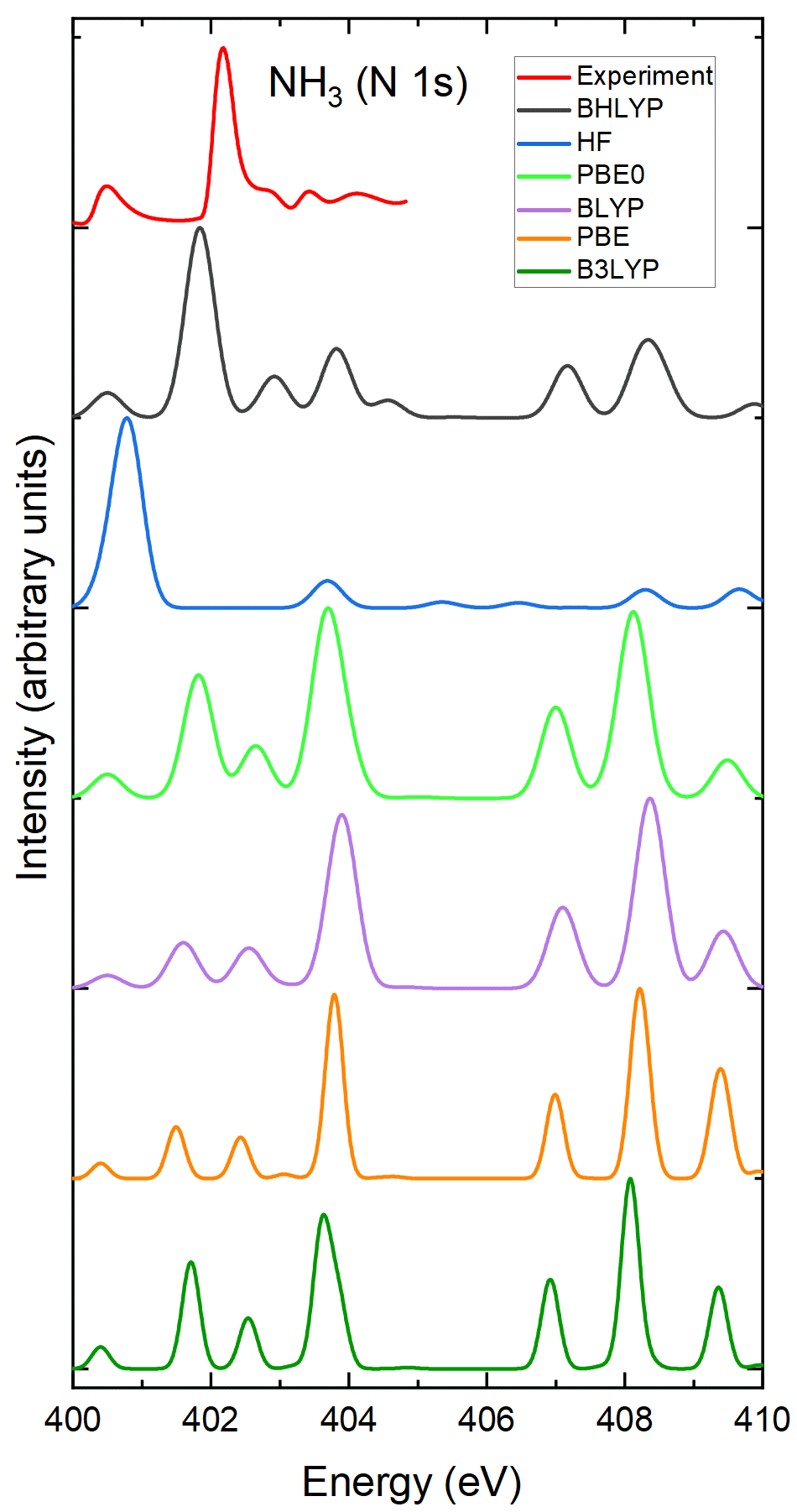}
    \caption{Comparison of measured XAS spectrum from Ref.~\cite{Nilsson2010b} and TDDFT results with different exchange-correlation functionals for NH$_3$. The calculated spectra were shifted such that the energy of the first excited state agrees with the measured value.}
    \label{func_comp}
\end{figure}

We have also studied the dependence of the TDDFT spectra on the basis set and the exchange-correlation functional. Fig.~\ref{basis_comp} compares the calculated spectra of the NH$_3$ molecule obtained using the BHLYP functional for different basis sets ranging from double zeta to quadruple zeta of the augmented correlation-consistent polarized, valence aug-cc-pVXZ Dunning family with X=D, T or Q~\cite{dunning1989a,kendall1992a} taken from the basis set exchange~\cite{pritchard2019a,feller1996a,schuchardt2007a}. In addition, we tested the aug-cc-pCVQZ basis which contains additional core basis functions. For the energy range in which experimental data is available, all basis sets give similar results. Clear differences can be observed at higher excitation energies. Finally, we also compare spectra with and without the Tamm-Dancoff approximation and found almost no difference. For all TDDFT calculations in this paper we use the aug-cc-pVQZ basis and the Tamm-Dancoff approximation.

Figure~\ref{func_comp} compares the experimental spectrum of NH$_3$ to calculated spectra obtained from different exchange-correlation functionals. Good agreement is found between the BHLYP result and experiment, while significant qualitative differences can be observed for the other functionals. All TDDFT calculations in the paper are therefore performed with the BHLYP functional.

\section{Results and Discussion}

\begin{figure}[h!]
    \centering
    \includegraphics[width=15cm]{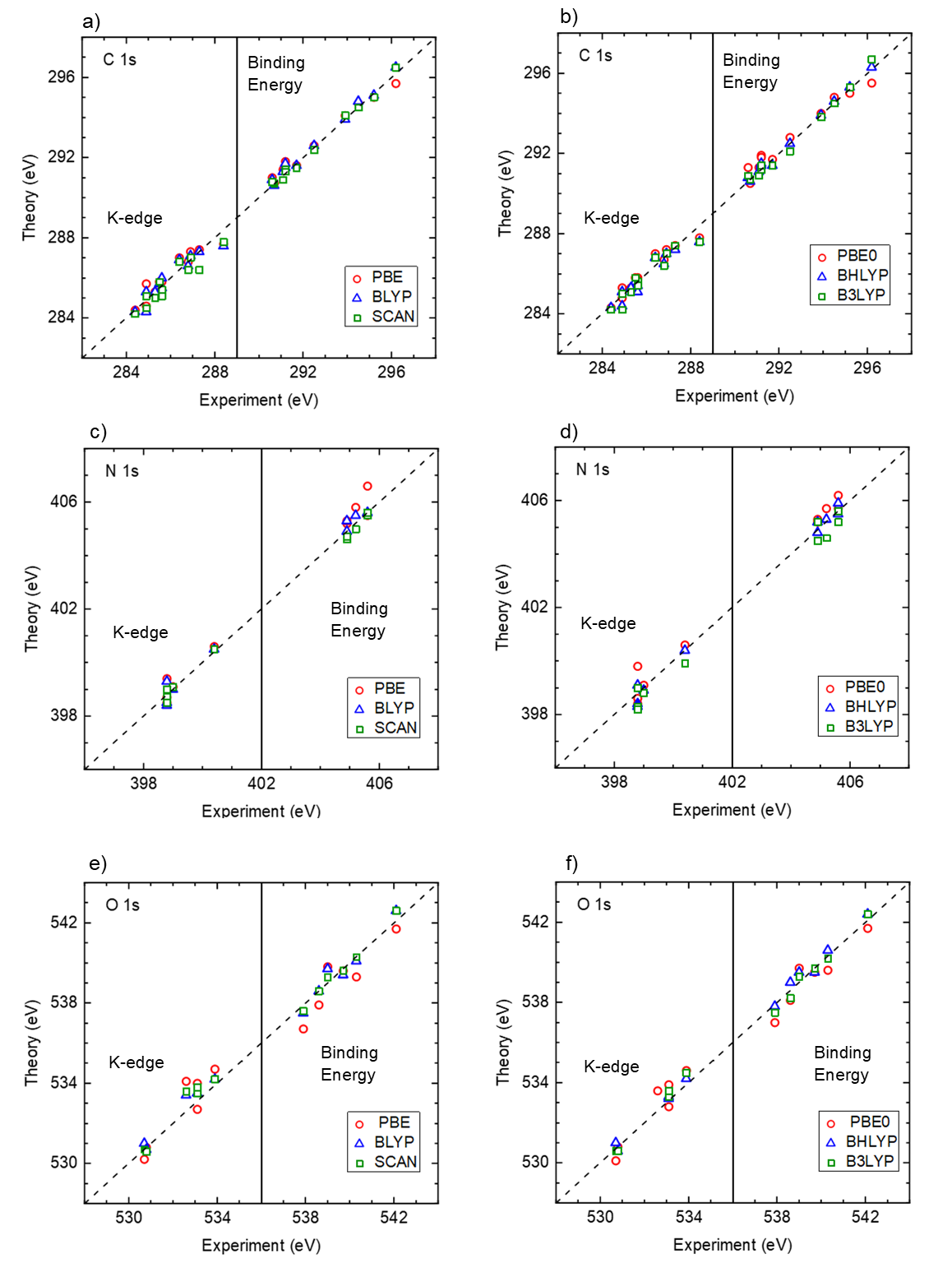}
    \caption{Comparison of calculated and measured core-electron binding energies (BEs) and K-edge energies. Results were obtained using the $\Delta$SCF approach with different exchange correlation functionals. The panels in the left column compare the performance of semi-local functionals (PBE, BLYP and SCAN) and the panels in the right column compare three different hybrid functionals (PBE0, BHLYP and B3LYP).}
    \label{fig:dks_graphs}
\end{figure}

\subsection{K-edge energies of molecular compounds}

We have calculated the lowest core-electron excitations energies corresponding to transitions from atomic 1s orbitals to molecular LUMO states (also known as the K-edge energy) as well as the core-electron binding energies of a set of small molecules containing the elements H, C, N, O and F using the $\Delta$SCF approach. In particular, we carry out calculations for \chem{CH_4}, \chem{H_2O}, \chem{NH_3}, HF, ethanol, acetone, CO, OCS, formaldehyde, \chem{C_2H_2}, \chem{C_2H_4} and azabenzenes. All results are shown in Figure~\ref{fig:dks_graphs} and summarized in Table~\ref{table3}. All relevant data used to generate these graphs as well as the experimental references is also provided in the Supplementary Materials.  

Figures~\ref{fig:dks_graphs}(a) and (b) compare the K-edge energies and core-electron binding energies of carbon atoms in the molecular compounds obtained with different exchange-correlation functionals. Table~\ref{table3} shows that the BHLYP functional performs best for the core-electron binding energies with a mean absolute error (MAE) of 0.11 eV, while the other hybrid functionals perform somewhat worse. Regarding the semi-local functionals, SCAN performs best with a MAE of 0.15 eV, while PBE and BLYP have MAEs of 0.22 eV and 0.29 eV, respectively. 
For the K-edge energies, PBE0 performs best with a MAE of 0.24 eV. B3LYP and BLYP (both with a MAE of 0.29 eV) as well as BHLYP (0.29 eV) and PBE (0.30 eV) perform similarly with SCAN showing the highest MAE of 0.38 eV. The origin of this large MAE for SCAN can be traced to its performance for carbon monoxide with an absolute error of 0.9 eV, while other functionals have errors of only 0.1 eV for this system. Overall, we find that the MAEs for the lowest neutral excitations tend to be somewhat higher than for the core-electron binding energies.

Figures~\ref{fig:dks_graphs}(c) and (d) show the corresponding results for oxygen atoms. Table~\ref{table3} shows that among the semi-local functionals, SCAN performs best for the core-electron binding energies with a MAE of 0.21 eV. For the K-edge energies, the accuracy of SCAN is somewhat worse with a MAE of 0.32 eV which is similar to BLYP with an MAE of 0.30 eV. Somewhat better results for the K-edge energies can be obtained with hybrid functionals. In particular, BHLYP yields a MAE of only 0.23 eV, while the MAEs of B3LYP and PBE0 are 0.32 eV and 0.48 eV, respectively.

Finally, Figs.~~\ref{fig:dks_graphs}(e) and (f) show the corresponding results for nitrogen atoms. For the core-electron binding energies, BLYP (MAE of 0.14 eV), SCAN (MAE of 0.16 eV), and BHLYP (MAE of 0.18 eV) perform best. For the K-edge energies, SCAN performs best with a MAE of 0.16 eV. BLYP, PBE and BHLYP all show MAEs of 0.26 eV. B3LYP and PBE0 perform worst with MAEs of 0.4 eV and 0.38 eV, respectively.

\begin{table}[h!]
\centering
\begin{tabular}{c|| m{1.2cm}| m{1.2cm}| m{1.2cm} | m{1.2cm} | m{1.2cm} | m{1.2cm} | m{1.2cm} | m{1.2cm} | m{1.2cm} | m{1.2cm} | m{1.2cm} | m{1.2cm}|} 
 \hline
 Element  & \multicolumn{2}{|c|}{B3LYP} & \multicolumn{2}{|c|}{BHLYP} & \multicolumn{2}{|c|}{BLYP} & \multicolumn{2}{|c|}{PBE0} & \multicolumn{2}{|c|}{PBE} & \multicolumn{2}{|c|}{SCAN} \\
 \hline
  & BE & K-edge & BE & K-edge & BE & K-edge & BE & K-edge & BE & K-edge & BE & K-edge \\
  \hline
 C & 0.18 & 0.29 & 0.11 & 0.28 & 0.22 & 0.29 & 0.40 & 0.24 & 0.29 & 0.30 & 0.15 & 0.38 \\
 O & 0.23 & 0.32 & 0.25 & 0.23 & 0.31 & 0.30 & 0.50 & 0.48 & 0.62 & 0.52 & 0.21 & 0.32 \\
 N & 0.25 & 0.40 & 0.18 & 0.26 & 0.14 & 0.26 & 0.38 & 0.38 & 0.48 & 0.26 & 0.16 & 0.16 \\
 F & 0.10 & 0.10 & 0.30 & 0.10 & 0.10 & 0.30 & 0.40 & 0.40 & 0.00 & 0.40 & 0.00 & 0.10 \\
 \hline
 average & 0.24 & 0.31 & 0.19 & 0.25 & 0.23 & 0.31 & 0.42 & 0.36 & 0.42 & 0.39 & 0.16 & 0.33
\end{tabular}
\caption{Performance of different exchange-correlation functionals for the calculations of core-electron binding energies (BE) and K-edge energies of molecular compounds containing the elements C, O, N and F. Shown are the Mean Absolute Errors (MAE) in eV. For the core-electron binding energy calculations the data sets consist of 9 (C), 6 (O), 5 (N) and 1 (F) different binding energies. For the K-edge energies, the data sets consist of 11 (C), 5 (O), 5 (N) and 1 (F) energies.}
\label{table3}
\end{table}

\subsection{Core-electron spectra of molecules}

\noindent
\begin{figure}[h!]
    {\includegraphics[width=13cm]{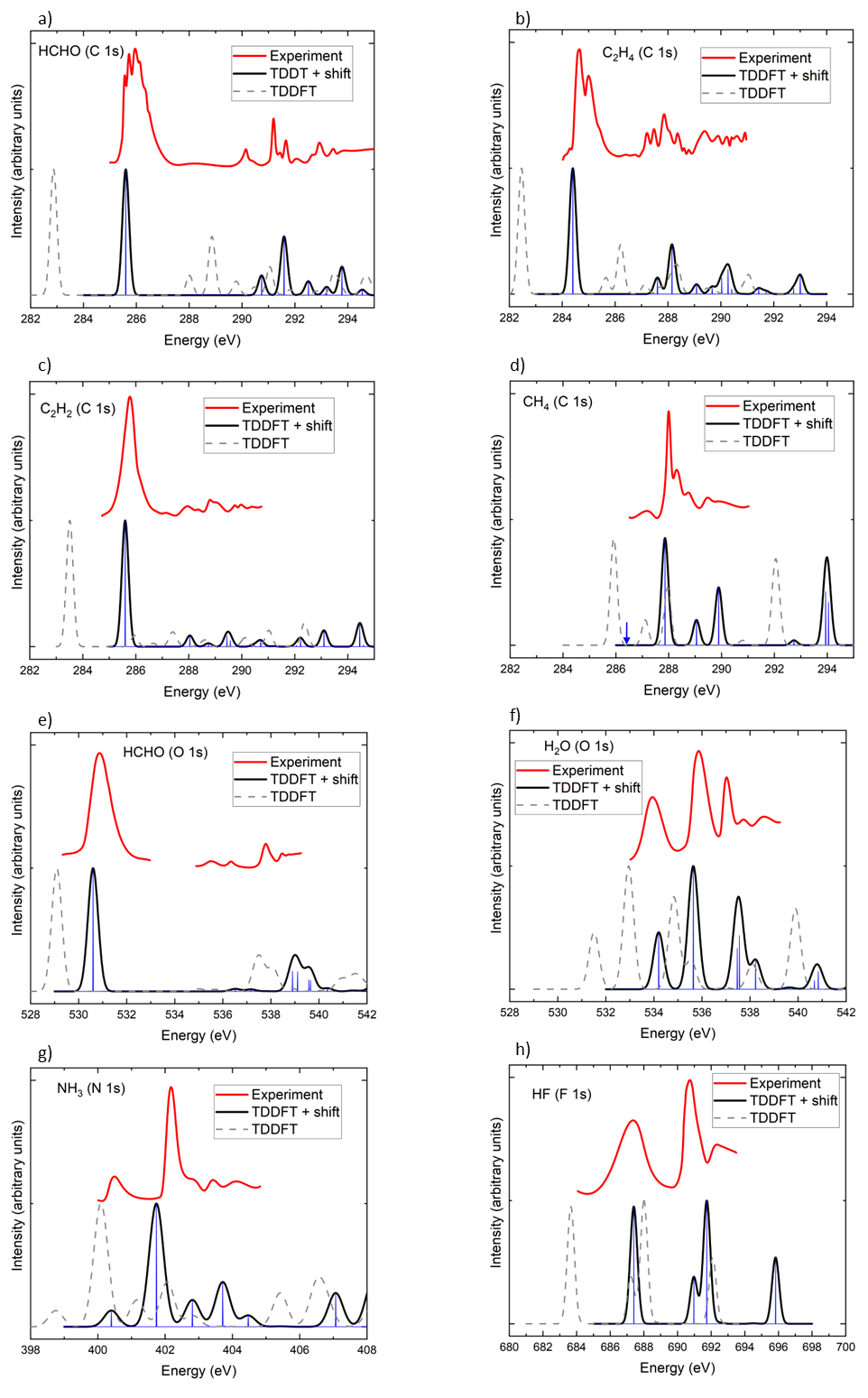}}
    \caption{Comparison of calculated and experimental core-electron spectra for a set of small molecules. The BHLYP TDDFT spectra without the $\Delta$SCF shift are shown as faint dashed lines. The blue arrow in the CH$_4$ spectrum indicates the position of the electric-dipole forbidden excitation which is visible in the experimental spectrum because of vibrational effects.}
    \label{fig:spectra_graphs}
\end{figure}

Figure~\ref{fig:spectra_graphs} shows the calculated core-level spectra for a set of small molecules (HCHO, C$_2$H$_2$, C$_2$H$_4$, CH$_4$, NH$_3$, H$_2$O, HF) and compares them to experimental results. The experimental data has been fitted to a spline and smoothed for easier visual comparison. As described in the methods section, each excitation is represented by a Lorentzian with a full width at half maximum $\eta$. In Fig.~\ref{fig:spectra_graphs}, we have used $\eta=0.3$~eV for carbon spectra and $0.5$~eV for oxygen, nitrogen and fluorine spectra. 

Figure~\ref{fig:spectra_graphs}(a) compares the calculated carbon 1s XAS spectrum of formaldehyde (HCHO) with experimental data taken from reference~\cite{Remmers1992a}. Both the experimental and the calculated spectrum show a large peak near 285.5 eV. In the measured spectrum, this peak appears very broad  and is split into a set of smaller subpeaks. This peak splitting has been interpreted as a vibrational effect~\cite{Remmers1992a} with a single electronic excitation (associated with the transition of an electron from the C 1s core level to an unoccupied $\pi^*$ orbital) coupled to various vibrations associated with H-C bending and stretching as well as C-O stretching. Since the atomic nuclei are fixed in our calculations, we do not capture these vibrational effects. Following this main peak, there is an energy gap in both the experimental and the calculated spectrum. At higher energies (starting at approximately 290 eV), a set of smaller peaks can be observed which arise from transitions from the C 1s state to Rydberg states of the molecule. Overall, there is good agreement both for the positions and intensities of the peaks between the calculated and the measured spectrum.  

The experimental core-electron spectrum of C$_2$H$_4$ (shown in Fig.~\ref{fig:spectra_graphs}(b)) is qualitatively similar to that of formaldehyde. In particular, a large peak is found at 284.7 eV which is split into two peaks because of the coupling to the symmetric C-H stretching mode. This peak arises from transitions from the carbon 1s orbital to the molecular LUMO. At energies higher than 287.7 eV, a series of smaller transitions is observed which are attributed to transitions into Rydberg states. The calculated spectrum also exhibits a large peak whose energy is in good agreement with the experimental one as well as a series of smaller peaks at higher energies. Similarly good agreement between theory and experiment is found for C$_2$H$_2$, see Fig.~\ref{fig:spectra_graphs}(c).

Figure~\ref{fig:spectra_graphs}(d) compares the XAS spectrum~\cite{Nilsson2010b} of CH$_4$ to the calculated result. In this case the agreement between theory and experiment is clearly worse. It is important to note, however, that the first low-intensity peak in the experimental spectrum at 287.05 eV arises from a transition from the carbon 1s orbital to 3s a$_1$ Rydberg orbital. This transition is electric-dipole forbidden and only observable because of vibrational coupling. The largest peak at 288.0 eV arises from a transition into the 3p t$_2$ Rydberg state and is followed by smaller peaks arising from vibrational effects. At higher energies additional peaks arising from transitions into higher Rydberg states are observed. In the calculated spectrum, the energy of the largest peak is underestimated, but better agreement is found for the higher-lying Rydberg state transitions.  

Figures~\ref{fig:spectra_graphs}(e) and (f) show the measured~\cite{Nilsson2010b,Remmers1992a} and calculated oxygen spectra of H$_2$O and CH$_2$O. For H$_2$O, very good agreement between theory and experiment is found both for the peak positions and their intensities. The first two peaks arise from transitions from oxygen 1s to the 4a$_1$ LUMO and 2b$_2$ LUMO+1 orbitals, respectively, while the final state of the third peak is a Rydberg state. Vibrational effects are responsible for the large width of the peaks. For CH$_2$O, the measured spectrum consists of a large peaks at about 530.8 eV which is well reproduced by theory and a series of smaller peaks arising from transitions into Rydberg states which are captured by the calculations. In particular, the calculated spectrum also exhibits two small peaks near 537 eV which arise from transitions into 3s and 3p Rydberg states followed by two somewhat larger peaks near 539 eV corresponding to transitions into 4p and 5p Rydberg states. However, the energies of these smaller peaks are approximately 1 eV higher in the calculated spectrum compared to experiment.

Finally, Fig.~\ref{fig:spectra_graphs}(g) and (h) show the nitrogen spectrum of NH$_3$ and the fluorine spectrum of HF, respectively, and compares them to experimental XAS results~\cite{Nilsson2010b}. Good agreement between theory and experiment is found for NH$_3$. In particular, the position and intensity of the first peak is well reproduced, but the energy of the large second peak near 402 eV is somewhat underestimated by the calculation. Similarly, the first peak at 687 eV in the HF spectrum is captured accurately by the calculation. At higher energies, near 692 eV, the theoretical spectra exhibits two peaks. In contrast to experiment, however, the intensity of the first peak is higher than that of the second peak.

\section{Conclusions}
We have assessed the performance of a first-principles approach for calculating core-electron spectra which are measured in X-ray absorption spectroscopy and energetic electron loss spectroscopy. In this approach, spectra from linear response TDDFT are shifted such that the energy of the lowest excitation agrees with the value obtained from $\Delta$SCF. This procedure overcomes TDDFT's failure to yield accurate absolute core-electron excitation energies, while producing the entire spectrum in one shot (as opposed to having a separate calculation for each excited state). We apply this method to set of small molecules and find mostly good agreement between experimental and calculated spectra when the BHLYP exchange-correlation functional is used for the TDDFT. This method can now be applied to more complex systems, including solids and surfaces.   

\section{Acknowledgements}
JL acknowledges funding from the Royal Society through a Royal Society University Research Fellowship. This project has received funding from the European Union's Horizon 2020 research and innovation programme under grant agreement No 892943.

\section{References}
\bibliographystyle{unsrt}
\bibliography{main.bib}

\section{appendix}

Table~\ref{table_1} shows $\Delta$SCF results for the binding energies for all molecules investigated in this study and Table~\ref{table_2} shows the K-edge energies. Experimental values are taken from: Ref.~\cite{Nilsson2010b} for \chem{CH_4}, \chem{H_2O}, \chem{NH_3}, HF; Ref.~\cite{Sham1989b} for ethanol, acetone, CO, OCS; Ref.~\cite{Remmers1992a} for formaldehyde; Ref.~\cite{Tronc1979a} for \chem{C_2H_2} and \chem{C_2H_4}; Ref.~\cite{Vall-Llosera2008b} for pyridine, pyridazine, pyrimidine and pyrazine.

\begin{table}[ht]
\centering
\caption{Core electron binding energies of molecules obtained from $\Delta$SCF. All energies are given in eV.}
\label{table_1}
\begin{tabular}{|l|l|r|c|r|c|c|c|c|}
\hline
Molecule & Atom & Exp. & B3LYP & BHLYP & BLYP & PBE0 & PBE & SCAN \\ \hline
CH$_3$ & C & 290.7 & 290.7 & 290.6 & 290.6 & 290.5 & 290.7 & 290.7 \\ \hline
NH$_3$ & N & 405.6 & 405.6 & 405.9 & 405.6 & 405.7 & 405.5 & 405.5 \\ \hline
H$_2$O & O & 539.7 & 539.7 & 539.5 & 539.4 & 539.5 & 539.6 & 539.6 \\ \hline
HF & F & 694.1 & 694.2 & 694.4 & 694.2 & 694.5 & 694.1 & 694.1 \\ \hline
OCS & C & 295.2 & 295.3 & 295.3 & 295.1 & 295.0 & 295.0 & 295.0 \\ \hline
 & O & 540.3 & \multicolumn{1}{r|}{540.2} & 540.6 & \multicolumn{1}{r|}{540.1} & \multicolumn{1}{r|}{539.6} & \multicolumn{1}{r|}{539.3} & \multicolumn{1}{r|}{540.3} \\ \hline
CO & C & 296.2 & 296.7 & 296.3 & 296.5 & 295.5 & 295.7 & 296.5 \\ \hline
 & O & 542.1 & 542.4 & 542.4 & 542.6 & 541.7 & 541.7 & 542.6 \\ \hline
Acetone & C(H$_3$) & 291.2 & 291.2 & 291.4 & 291.5 & 291.9 & 291.8 & 291.4 \\ \hline
 & C(O) & 293.9 & 293.8 & 293.9 & 293.9 & 294.0 & 294.1 & 294.1 \\ \hline
 & O & 537.9 & \multicolumn{1}{r|}{537.5} & 537.8 & \multicolumn{1}{r|}{537.5} & \multicolumn{1}{r|}{537.0} & \multicolumn{1}{r|}{536.7} & \multicolumn{1}{r|}{537.6} \\ \hline
C$_2$H$_4$ & C & 290.6 & 290.9 & 290.8 & 290.9 & 291.3 & 291.0 & 290.8 \\ \hline
C$_2$H$_2$ & C & 291.2 & 291.4 & 291.5 & 291.7 & 291.8 & 291.6 & 291.3 \\ \hline
Formaldehyde & C & 294.5 & 294.5 & 294.6 & 294.8 & 294.8 & 294.7 & \multicolumn{1}{r|}{294.5} \\ \hline
 &  & 539.0 & 539.3 & 539.5 & 539.7 & 539.7 & 539.8 & \multicolumn{1}{r|}{539.3} \\ \hline
Ethanol & C(H$_2$OH) & 291.1 & 290.9 & 291.1 & 291.3 & 291.3 & 291.4 & 290.9 \\ \hline
 & C(H$_3$­) & 292.5 & 292.1 & 292.5 & 292.6 & 292.8 & 292.6 & 292.4 \\ \hline
 & O & 538.6 & 538.2 & 539.0 & \multicolumn{1}{r|}{538.6} & \multicolumn{1}{r|}{538.1} & \multicolumn{1}{r|}{537.9} & \multicolumn{1}{r|}{538.6} \\ \hline
Pyridine & N & 404.9 & 404.5 & 404.8 & 404.9 & 405.2 & 405.2 & 404.6 \\ \hline
Pyridazine & N & 404.9 & 405.2 & 405.2 & 405.3 & 405.3 & 405.3 & 404.7 \\ \hline
Pyrimidine & N & 405.2 & 404.6 & 405.3 & 405.5 & 405.7 & 405.8 & 405.0 \\ \hline
Pyrazine & C & 291.7 & 291.4 & 291.4 & 291.6 & 291.7 & 291.6 & 291.5 \\ \hline
 & N & 405.6 & 405.2 & 405.5 & 405.6 & 406.2 & 406.6 & 405.6 \\ \hline \hline
\textbf{MAE} &  &  & \multicolumn{1}{r|}{0.2} & 0.2 & \multicolumn{1}{r|}{0.2} & \multicolumn{1}{r|}{0.4} & \multicolumn{1}{r|}{0.4} & \multicolumn{1}{r|}{0.2} \\ \hline
\end{tabular}%
\end{table}

\begin{table}[ht]
\centering
\caption{K-edge energies of molecules obtained from $\Delta$SCF. All energies are given in eV. For excitations from O 1s in ethanol, results could not be obtained for B3LYP and BHLYP because of variational collapse problems (indicated by ***).
}
\label{table_2}
\begin{tabular}{|
>{\columncolor[HTML]{FFFFFF}}l |
>{\columncolor[HTML]{FFFFFF}}l |
>{\columncolor[HTML]{FFFFFF}}c |
>{\columncolor[HTML]{FFFFFF}}c |
>{\columncolor[HTML]{FFFFFF}}c |
>{\columncolor[HTML]{FFFFFF}}c |
>{\columncolor[HTML]{FFFFFF}}c |
>{\columncolor[HTML]{FFFFFF}}c |
>{\columncolor[HTML]{FFFFFF}}c |}
\hline
{\color[HTML]{333333} Molecule} & {\color[HTML]{333333} Atom} & \multicolumn{1}{r|}{\cellcolor[HTML]{FFFFFF}{\color[HTML]{333333} Experiment}} & {\color[HTML]{333333} B3LYP} & \multicolumn{1}{r|}{\cellcolor[HTML]{FFFFFF}{\color[HTML]{333333} BHLYP}} & {\color[HTML]{333333} BLYP} & {\color[HTML]{333333} PBE0} & {\color[HTML]{333333} PBE} & {\color[HTML]{333333} SCAN} \\ \hline
{\color[HTML]{333333} CH$_3$} & {\color[HTML]{333333} C} & {\color[HTML]{333333} 286.8} & {\color[HTML]{333333} 286.4} & {\color[HTML]{333333} 286.5} & {\color[HTML]{333333} 286.6} & {\color[HTML]{333333} 286.7} & {\color[HTML]{333333} 286.8} & {\color[HTML]{333333} 286.4} \\ \hline
{\color[HTML]{333333} NH$_3$} & {\color[HTML]{333333} N} & {\color[HTML]{333333} 400.4} & {\color[HTML]{333333} 399.9} & {\color[HTML]{333333} 400.4} & {\color[HTML]{333333} 400.5} & {\color[HTML]{333333} 400.6} & {\color[HTML]{333333} 400.6} & {\color[HTML]{333333} 400.5} \\ \hline
{\color[HTML]{333333} H$_2$O} & {\color[HTML]{333333} O} & {\color[HTML]{333333} 533.9} & {\color[HTML]{333333} 534.5} & {\color[HTML]{333333} 534.2} & {\color[HTML]{333333} 534.2} & {\color[HTML]{333333} 534.6} & {\color[HTML]{333333} 534.7} & {\color[HTML]{333333} 534.2} \\ \hline
{\color[HTML]{333333} HF} & {\color[HTML]{333333} F} & {\color[HTML]{333333} 687.3} & {\color[HTML]{333333} 687.4} & {\color[HTML]{333333} 687.4} & {\color[HTML]{333333} 687.6} & {\color[HTML]{333333} 687.7} & {\color[HTML]{333333} 687.7} & {\color[HTML]{333333} 687.4} \\ \hline
{\color[HTML]{333333} OCS} & {\color[HTML]{333333} C} & {\color[HTML]{333333} 288.4} & {\color[HTML]{333333} 287.6} & {\color[HTML]{333333} 287.6} & {\color[HTML]{333333} 287.6} & {\color[HTML]{333333} 287.8} & {\color[HTML]{333333} 287.7} & {\color[HTML]{333333} 287.8} \\ \hline
{\color[HTML]{333333} } & {\color[HTML]{333333} O} & {\color[HTML]{333333} 533.1} & {\color[HTML]{333333} 533.3} & {\color[HTML]{333333} 533.2} & {\color[HTML]{333333} 533.5} & {\color[HTML]{333333} 532.8} & {\color[HTML]{333333} 532.7} & {\color[HTML]{333333} 533.5} \\ \hline
{\color[HTML]{333333} CO} & {\color[HTML]{333333} C} & {\color[HTML]{333333} 287.3} & {\color[HTML]{333333} 287.4} & {\color[HTML]{333333} 287.2} & {\color[HTML]{333333} 287.3} & {\color[HTML]{333333} 287.4} & {\color[HTML]{333333} 287.4} & {\color[HTML]{333333} 286.4} \\ \hline
{\color[HTML]{333333} } & {\color[HTML]{333333} O} & {\color[HTML]{333333} 533.1} & {\color[HTML]{333333} 533.6} & {\color[HTML]{333333} 533.3} & {\color[HTML]{333333} 533.5} & {\color[HTML]{333333} 533.9} & {\color[HTML]{333333} 534.0} & {\color[HTML]{333333} 533.8} \\ \hline
{\color[HTML]{333333} Acetone} & {\color[HTML]{333333} C(H$_3$)} & {\color[HTML]{333333} 286.4} & {\color[HTML]{333333} 286.8} & {\color[HTML]{333333} 286.8} & {\color[HTML]{333333} 286.9} & {\color[HTML]{333333} 287.0} & {\color[HTML]{333333} 287.0} & {\color[HTML]{333333} 286.8} \\ \hline
{\color[HTML]{333333} } & {\color[HTML]{333333} O} & {\color[HTML]{333333} 530.7} & {\color[HTML]{333333} 530.6} & {\color[HTML]{333333} 531.0} & {\color[HTML]{333333} 531.0} & {\color[HTML]{333333} 530.1} & {\color[HTML]{333333} 530.2} & {\color[HTML]{333333} 530.7} \\ \hline
{\color[HTML]{333333} C$_2$H$_4$} & {\color[HTML]{333333} C} & {\color[HTML]{333333} 284.4} & {\color[HTML]{333333} 284.2} & {\color[HTML]{333333} 284.3} & {\color[HTML]{333333} 284.3} & {\color[HTML]{333333} 284.3} & {\color[HTML]{333333} 284.4} & {\color[HTML]{333333} 284.2} \\ \hline
{\color[HTML]{333333} C$_2$H$_2$} & {\color[HTML]{333333} C} & {\color[HTML]{333333} 285.6} & {\color[HTML]{333333} 285.6} & {\color[HTML]{333333} 285.1} & {\color[HTML]{333333} 286.0} & {\color[HTML]{333333} 285.8} & {\color[HTML]{333333} 285.4} & {\color[HTML]{333333} 285.1} \\ \hline
{\color[HTML]{333333} Formaldehyde} & {\color[HTML]{333333} C} & {\color[HTML]{333333} 285.6} & {\color[HTML]{333333} 285.4} & {\color[HTML]{333333} 285.4} & {\color[HTML]{333333} 285.5} & {\color[HTML]{333333} 285.7} & {\color[HTML]{333333} 285.8} & {\color[HTML]{333333} 285.4} \\ \hline
{\color[HTML]{333333} } & {\color[HTML]{333333} O} & {\color[HTML]{333333} 530.8} & {\color[HTML]{333333} 530.6} & {\color[HTML]{333333} 530.6} & {\color[HTML]{333333} 530.7} & {\color[HTML]{333333} 530.8} & {\color[HTML]{333333} 530.8} & {\color[HTML]{333333} 530.6} \\ \hline
{\color[HTML]{333333} Ethanol} & {\color[HTML]{333333} C(H$_2$OH)} & {\color[HTML]{333333} 286.9} & {\color[HTML]{333333} 287.0} & {\color[HTML]{333333} 287.0} & {\color[HTML]{333333} 287.1} & {\color[HTML]{333333} 287.2} & {\color[HTML]{333333} 287.3} & {\color[HTML]{333333} 287.0} \\ \hline
{\color[HTML]{333333} } & {\color[HTML]{333333} O} & {\color[HTML]{333333} 532.6} & {\color[HTML]{333333} ***} & \multicolumn{1}{l|}{\cellcolor[HTML]{FFFFFF}{\color[HTML]{333333} ***}} & \multicolumn{1}{l|}{\cellcolor[HTML]{FFFFFF}{\color[HTML]{333333} 533.4}} & {\color[HTML]{333333} 533.6} & {\color[HTML]{333333} 534.1} & {\color[HTML]{333333} 533.6} \\ \hline
{\color[HTML]{333333} Pyridine} & {\color[HTML]{333333} C} & {\color[HTML]{333333} 284.9} & {\color[HTML]{333333} 284.2} & {\color[HTML]{333333} 284.4} & {\color[HTML]{333333} 284.3} & {\color[HTML]{333333} 284.8} & {\color[HTML]{333333} 284.6} & {\color[HTML]{333333} 284.5} \\ \hline
{\color[HTML]{333333} } & {\color[HTML]{333333} N} & {\color[HTML]{333333} 398.8} & {\color[HTML]{333333} 398.3} & {\color[HTML]{333333} 398.4} & {\color[HTML]{333333} 398.5} & {\color[HTML]{333333} 398.6} & {\color[HTML]{333333} 398.7} & {\color[HTML]{333333} 398.7} \\ \hline
{\color[HTML]{333333} Pyridazine} & {\color[HTML]{333333} C} & {\color[HTML]{333333} 285.5} & {\color[HTML]{333333} 285.8} & {\color[HTML]{333333} 285.6} & {\color[HTML]{333333} 285.7} & {\color[HTML]{333333} 285.8} & {\color[HTML]{333333} 285.8} & {\color[HTML]{333333} 285.8} \\ \hline
{\color[HTML]{333333} } & {\color[HTML]{333333} N} & {\color[HTML]{333333} 399.0} & {\color[HTML]{333333} 398.8} & {\color[HTML]{333333} 398.9} & {\color[HTML]{333333} 399.0} & {\color[HTML]{333333} 399.1} & {\color[HTML]{333333} 399.1} & {\color[HTML]{333333} 399.1} \\ \hline
{\color[HTML]{333333} Pyrimidine} & {\color[HTML]{333333} C} & {\color[HTML]{333333} 284.9} & {\color[HTML]{333333} 285.0} & {\color[HTML]{333333} 285.1} & {\color[HTML]{333333} 285.3} & {\color[HTML]{333333} 285.3} & {\color[HTML]{333333} 285.7} & {\color[HTML]{333333} 285.1} \\ \hline
{\color[HTML]{333333} } & {\color[HTML]{333333} N} & {\color[HTML]{333333} 398.8} & {\color[HTML]{333333} 398.2} & {\color[HTML]{333333} 398.3} & {\color[HTML]{333333} 398.4} & {\color[HTML]{333333} 398.4} & {\color[HTML]{333333} 398.5} & {\color[HTML]{333333} 398.5} \\ \hline
{\color[HTML]{333333} Pyrazine} & {\color[HTML]{333333} C} & {\color[HTML]{333333} 285.3} & {\color[HTML]{333333} 285.1} & {\color[HTML]{333333} 285.3} & {\color[HTML]{333333} 285.3} & {\color[HTML]{333333} 285.3} & {\color[HTML]{333333} 285.3} & {\color[HTML]{333333} 285.0} \\ \hline
{\color[HTML]{333333} } & {\color[HTML]{333333} N} & {\color[HTML]{333333} 398.8} & {\color[HTML]{333333} 399.0} & {\color[HTML]{333333} 399.1} & {\color[HTML]{333333} 399.3} & {\color[HTML]{333333} 399.8} & {\color[HTML]{333333} 399.4} & {\color[HTML]{333333} 399.0} \\ \hline \hline
{\color[HTML]{333333} \textbf{MAE}} & {\color[HTML]{333333} } & \multicolumn{1}{r|}{\cellcolor[HTML]{FFFFFF}{\color[HTML]{333333} }} & \multicolumn{1}{r|}{\cellcolor[HTML]{FFFFFF}{\color[HTML]{333333} 0.3}} & \multicolumn{1}{r|}{\cellcolor[HTML]{FFFFFF}{\color[HTML]{333333} 0.3}} & \multicolumn{1}{r|}{\cellcolor[HTML]{FFFFFF}{\color[HTML]{333333} 0.3}} & \multicolumn{1}{r|}{\cellcolor[HTML]{FFFFFF}{\color[HTML]{333333} 0.4}} & \multicolumn{1}{r|}{\cellcolor[HTML]{FFFFFF}{\color[HTML]{333333} 0.4}} & \multicolumn{1}{r|}{\cellcolor[HTML]{FFFFFF}{\color[HTML]{333333} 0.3}} \\ \hline
\end{tabular}%
\end{table}

\end{document}